\documentstyle[psfig,12pt]{article} 
\def\lsim{\:\raisebox{-0.5ex}{$\stackrel{\textstyle<}{\sim}$}\:} 
 
\def\be{\begin{equation}}        
\def\ee{\end{equation}} 
\def\bear{\be\begin{array}}       
\def\eear{\end{array}\ee} 
\def\bea{\begin{eqnarray}} 
\def\eea{\end{eqnarray}}

\def\21{$SU(2) \ot U(1)$} 
\def\ot{\otimes} 
\def\ie{{\it i.e.}} 
 
\def\half{{\textstyle{1 \over 2}}} 
\def\third{{\textstyle{1 \over 3}}} 
\def\quarter{{\textstyle{1 \over 4}}} 
 
\def\eighth{{\textstyle{1 \over 8}}} 
\def\bold#1{\setbox0=\hbox{$#1$} 
     \kern-.025em\copy0\kern-\wd0 
     \kern.05em\copy0\kern-\wd0 
     \kern-.025em\raise.0433em\box0 } 
\begin{document} 
\begin{titlepage} 
\begin{flushright} 
hep-ph/9707395 \\ 
FTUV/97-33\\ 
IFIC/97-32\\ 
July 1997 
\end{flushright} 
\vspace*{5mm} 
\begin{center}  
{\Large \bf Charged Higgs Boson and Stau Phenomenology in the 
Simplest R--Parity Breaking Model}\\[15mm] 
{\large A. Akeroyd, Marco A. D\'\i az, J. Ferrandis, 
M. A. Garc{\'\i}a-Jare\~no} \\ 
and \\ 
{\large Jos\'e W. F. Valle}\\ 
\hspace{3cm}\\ 
{\small Departamento de F\'\i sica Te\'orica, IFIC-CSIC, Universidad de Valencia}\\  
{\small Burjassot, Valencia 46100, Spain} 
\end{center} 
\vspace{5mm} 
\begin{abstract} 
 
We consider the charged scalar boson phenomenology in the simplest 
effective low-energy R-parity breaking model characterized by a 
bilinear violation of R-parity in the superpotential.  This induces a 
mixing between staus and the charged Higgs boson. We show that the 
charged Higgs boson mass can be lower than expected in the MSSM, even 
before including radiative corrections. We also study the charged 
scalar boson decay branching ratios and show that the R-parity 
violating decay rates can be comparable or even bigger than the 
R-parity conserving ones. Moreover, if the stau is the LSP it will 
have only decays into standard model fermions. These features could 
have important implications for charged supersymmetric scalar boson 
searches at future accelerators. 
 
\end{abstract} 
 
\end{titlepage} 
\setcounter{page}{1} 
 
\section{Introduction} 
 
A lot of emphasis has been put into the phenomenological study of the 
supersymmetric Higgs boson sector \cite{carena}. However, so far most 
of these phenomenological studies have been made in the framework of 
the Minimal Supersymmetric Standard Model (MSSM) \cite{mssm,HabKane} with 
conserved R-parity \cite{RP}.  R-parity is a discrete symmetry 
assigned as $R_p=(-1)^{(3B+L+2S)}$, where L is the lepton number, B is 
the baryon number and S is the spin of the state. If R-parity is 
conserved all supersymmetric particles must always be pair-produced, 
while the lightest of them must be stable.  In particular, 
supersymmetric Higgs scalar bosons must decay only to normal standard 
model particles or to pairs of lowest-lying supersymmetric particles, 
which are usually heavy. On the other hand staus decay only to 
supersymmetric states, like a neutralino and a tau lepton. 
 
The study of alternative supersymmetric scenarios where the effective 
low energy theory violates R-parity \cite{HallSuzuki} has recently 
received a lot of attention \cite{beyond} both due to its 
phenomenological interest, as well as due to the intrinsic importance 
of investigating the issue of R-parity breaking at a deeper level. 
 
It is well--known that the simplest supersymmetric extension of the 
Standard Model violates R-parity through a set of cubic superpotential 
terms involving a very large number of arbitrary Yukawa 
couplings. Although highly constrained by proton stability, many of 
such scenarios could still be viable. Nevertheless their systematic 
study at a phenomenological level is hardly possible, due to the 
enormous number of parameters present, in addition to those of the 
MSSM. 
 
As with other fundamental symmetries, it could well be that R-parity is a 
symmetry at the Lagrangian level but is broken by the ground state. 
In order to comply with LEP precision measurements of the invisible Z 
decay width these models require the introduction of \21 singlet 
superfields \cite{mv90}. Such scenarios provide a very {\sl 
systematic} way to include R parity violating effects, automatically 
consistent with low energy {\sl baryon number conservation}. They have 
many added virtues, such as the possibility of providing a dynamical 
origin for the breaking of R-parity, through radiative corrections, 
similar to the electroweak symmetry \cite{rprad}.  The simplest 
truncated version of such a model, in which the violation of R-parity 
is effectively parametrized by a bilinear superpotential term 
$\epsilon_i\widehat L_i^a\widehat H_2^b$ has been widely discussed 
\cite{RPeps,epsrad}. It has also been shown recently \cite{epsrad} that this 
model is consistent with minimal N=1 supergravity unification with 
radiative breaking of the electroweak symmetry and universal scalar 
and gaugino masses. This one-parameter extension of the MSSM-SUGRA 
model therefore provides the simplest reference model for the breaking 
of R-parity and constitutes a consistent truncation of the complete 
dynamical models with spontaneous R-parity breaking proposed 
previously \cite{mv90}.  In this case there is no physical Goldstone 
boson, the Majoron, associated to the spontaneous breaking of 
R-parity, since in this effective truncated model the superfield 
content is exactly the standard one of the MSSM. Formulated as an 
effective theory at the weak scale, the model contains only two new 
parameters in addition to those of the MSSM. Therefore our model 
provides also the simplest parametrization of R-parity breaking 
effects.  In contrast to models with tri-linear R-parity breaking 
couplings, it leads to a very restrictive and systematic pattern of 
R-parity violating interactions, which can be taken as a reference 
model. 
 
In this paper we focus on the phenomenology of the charged scalar 
boson sector of the simplest R-parity breaking model. This complements 
a previous study of the electrically neutral sector \cite{eps0}. We 
show that 
\begin{enumerate} 
\item 
the mass of the charged Higgs boson can be lower than expected in the 
MSSM, even before including radiative corrections, 
\item 
if the stau is the LSP it will have only R-parity violating decay 
channels into standard model fermions, 
\item 
the branching ratio for the R-parity violating charged Higgs boson 
decays can be comparable or even bigger than the R-parity conserving 
ones. 
\end{enumerate} 
We illustrate how these features arising from the charged scalar boson 
sector of the simplest R-parity breaking model could play an important 
role in designing charged supersymmetric scalar boson searches at $e^+ 
e^-$ colliders such as LEP II. For example they can give rise to 
striking signatures consisting of high multiplicity events, such as 
di--tau + 4 jets + missing energy or 4 taus + 4 jets. Such processes, 
forbidden in the MSSM, are expected to have high rates and negligible 
background.  As for hadron colliders, we also can have very high 
leptonic multiplicity events such as six leptons of which at least two 
are taus, plus missing momentum. This should be easy to see at the 
LHC, due again to the negligible standard model background. 
 
\section{The Model} 
 
The supersymmetric Lagrangian is specified by the superpotential $W$ 
given by 
\footnote{We are using here the notation of refs.~\cite{HabKane} and 
\cite{GunHaber}.  }  
\begin{equation}  
W=\varepsilon_{ab}\left[ 
 h_U^{ij}\widehat Q_i^a\widehat U_j\widehat H_2^b 
+h_D^{ij}\widehat Q_i^b\widehat D_j\widehat H_1^a 
+h_E^{ij}\widehat L_i^b\widehat R_j\widehat H_1^a 
-\mu\widehat H_1^a\widehat H_2^b 
+\epsilon_i\widehat L_i^a\widehat H_2^b\right] 
\label{eq:Wsuppot} 
\end{equation} 
where $i,j=1,2,3$ are generation indices, $a,b=1,2$ are $SU(2)$ 
indices, and $\varepsilon$ is a completely antisymmetric $2\times2$  
matrix, with $\varepsilon_{12}=1$. The symbol ``hat'' over each  
letter indicates a superfield, with $\widehat Q_i$, $\widehat L_i$,  
$\widehat H_1$, and $\widehat H_2$ being $SU(2)$ doublets with  
hypercharges $\third$, $-1$, $-1$, and $1$ respectively, and $\widehat 
U$, $\widehat D$, and $\widehat R$ 
being $SU(2)$ singlets with hypercharges $-{\textstyle{4\over 3}}$, 
${\textstyle{2\over 3}}$, and $2$ respectively. The couplings $h_U$, 
$h_D$ and $h_E$ are $3\times 3$ Yukawa matrices, and $\mu$ and $\epsilon_i$ 
are parameters with units of mass. The last term in  
eq.~(\ref{eq:Wsuppot}) is the only $R$--parity violating term. 
 
Supersymmetry breaking is parametrized with a set of soft 
supersymmetry breaking terms which do not introduce quadratic 
divergences to the unrenormalized theory \cite{softterms} 
\begin{eqnarray} 
V_{soft}&=& 
M_Q^{ij2}\widetilde Q^{a*}_i\widetilde Q^a_j+M_U^{ij2} 
\widetilde U^*_i\widetilde U_j+M_D^{ij2}\widetilde D^*_i 
\widetilde D_j+M_L^{ij2}\widetilde L^{a*}_i\widetilde L^a_j+ 
M_R^{ij2}\widetilde R^*_i\widetilde R_j \nonumber\\ 
&&\!\!\!\!+m_{H_1}^2 H^{a*}_1 H^a_1+m_{H_2}^2 H^{a*}_2 H^a_2- 
\left[\half M_s\lambda_s\lambda_s+\half M\lambda\lambda 
+\half M'\lambda'\lambda'+h.c.\right]\label{eq:Vsoft} \\ 
&&\!\!\!\!\!\!\!\!\!\!\!\!\!\!\!\!\!\!\!\!+\varepsilon_{ab}\left[ 
A_U^{ij}h_U^{ij}\widetilde Q_i^a\widetilde U_j H_2^b 
+A_D^{ij}h_D^{ij}\widetilde Q_i^b\widetilde D_j H_1^a 
+A_E^{ij}h_E^{ij}\widetilde L_i^b\widetilde R_j H_1^a 
-B\mu H_1^a H_2^b+B_2\epsilon_i\widetilde L_i^a H_2^b\right] 
\,,\nonumber 
\end{eqnarray} 
and again, the last term in eq.~(\ref{eq:Vsoft}) is the only R--parity 
violating term. 
 
Following previous discussions we will focus for simplicity on the 
case of one generation, namely the third \cite{eps0,sensi}.   
In contrast we will keep in our discussion the theory as 
defined at low energies by the most general set of soft-breaking 
masses, tri-linear and bilinear  soft-breaking parameters, 
gaugino masses and the Higgs superfield mixing  parameter $\mu$. 
 
The electroweak symmetry is broken when the two Higgs doublets  
$H_1$ and $H_2$, and the third component of the left slepton 
doublet $\widetilde L_3$ acquire vacuum expectation values. We 
introduce the notation: 
\begin{eqnarray} 
&&H_1={{{1\over{\sqrt{2}}}[\chi^0_1+v_1+i\varphi^0_1]}\choose{ 
H^-_1}}\,,\qquad 
H_2={{H^+_2}\choose{{1\over{\sqrt{2}}}[\chi^0_2+v_2+ 
i\varphi^0_2]}}\,, 
\nonumber \\ 
&&\qquad\qquad\qquad\widetilde L_3={{{1\over{\sqrt{2}}} 
[\tilde\nu^R_{\tau}+v_3+i\tilde\nu^I_{\tau}]}\choose{\tilde\tau^-}}\,. 
\label{eq:shiftdoub} 
\end{eqnarray} 
Note that the $W$ boson acquires a mass $m_W^2=\quarter g^2v^2$, where 
$v^2\equiv v_1^2+v_2^2+v_3^2=(246 \; \rm{GeV})^2$. We introduce the 
following notation in spherical coordinates for the vacuum expectation  
values (VEV): 
\begin{eqnarray} 
v_1&=&v\sin\theta\cos\beta\cr 
v_2&=&v\sin\theta\sin\beta\cr 
v_3&=&v\cos\theta 
\label{eq:vevs} 
\end{eqnarray} 
which preserves the MSSM definition $\tan\beta=v_2/v_1$. In the MSSM limit, 
where $\epsilon_3=v_3=0$, the angle $\theta$ is equal to $\pi/2$.  
 
In addition to the above MSSM parameters, our model contains three new 
parameters, $\epsilon_3$, $v_3$ and $B_2$, of which only two are 
independent, and these may be chosen as $\epsilon_3$ and $v_3$.  
 
The full scalar potential at tree level is 
\begin{equation} 
V_{total}  = \sum_i \left| { \partial W \over \partial z_i} \right|^2 
	+ V_D + V_{soft} 
\label{V} 
\end{equation} 
where $z_i$ is any one of the scalar fields in the superpotential, 
$V_D$ are the $D$-terms, and $V_{soft}$ the SUSY soft 
breaking terms given in eq.~(\ref{eq:Vsoft}).  
 
The scalar potential contains linear terms 
\begin{equation} 
V_{linear}=t_1^0\chi^0_1+t_2^0\chi^0_2+t_3^0\tilde\nu^R_{\tau}\,, 
\label{eq:Vlinear} 
\end{equation} 
where 
\begin{eqnarray} 
t_1^0&=&(m_{H_1}^2+\mu^2)v_1-B\mu v_2-\mu\epsilon_3v_3+ 
\eighth(g^2+g'^2)v_1(v_1^2-v_2^2+v_3^2)\,, 
\nonumber \\ 
t_2^0&=&(m_{H_2}^2+\mu^2+\epsilon_3^2)v_2-B\mu v_1+B_2\epsilon_3v_3- 
\eighth(g^2+g'^2)v_2(v_1^2-v_2^2+v_3^2)\,, 
\label{eq:tadpoles} \\ 
t_3^0&=&(m_{L_3}^2+\epsilon_3^2)v_3-\mu\epsilon_3v_1+B_2\epsilon_3v_2+ 
\eighth(g^2+g'^2)v_3(v_1^2-v_2^2+v_3^2)\,. 
\nonumber 
\end{eqnarray} 
These $t_i^0, i=1,2,3$ are the tree level tadpoles, and are equal to  
zero at the minimum of the potential.  
 
Now a few theoretical comments on the model. The first refers to the
choice of basis in the original Lagrangian in eq.~(\ref{eq:Wsuppot}). We
could have used a rotated basis in which the bi-linear coupling
disappears \cite{dreiner}. However, if we choose to do so, new
R--parity violating terms appear not only in the tri-linear
superpotential, in the form of a $D Q L$ term, but also in the scalar
sector of the theory, i.e. the Higgs potential due to supersymmetry
breaking. Authors who adopt this basis \cite{sacha} have a tendency to
neglect SUSY breaking in the Higgs potential, which is most crucial
for our subsequent analysis. Therefore we prefer to use in our
calculations the basis in which the bi-linear term is not rotated
away.  The final physics results are completely basis-independent
\cite{DJV}. 

\begin{figure}
\centerline{\protect\hbox{\psfig{file=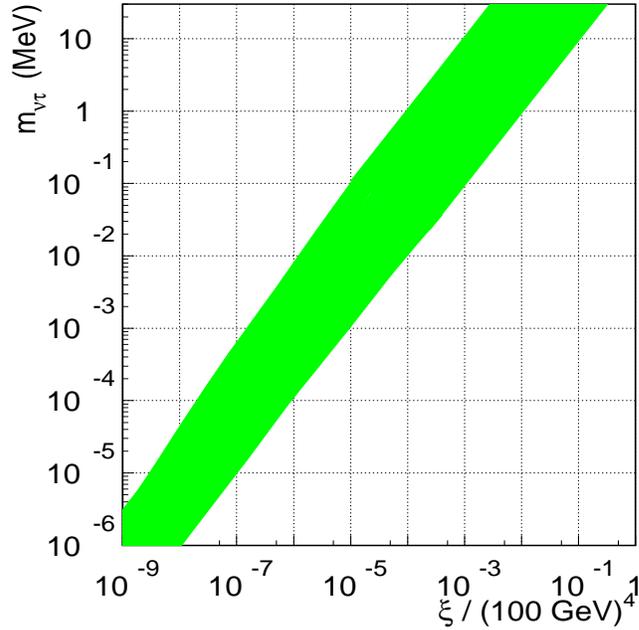,height=9cm,width=0.6\textwidth}}}
\caption{Tau neutrino mass versus de R--parity violating parameter $\xi$ }
\label{mnt_xi_new}
\end{figure}

Another important feature of this model is that lepton number is
violated by the $\epsilon_3$ term and by the presence of the sneutrino
vacuum expectation value $v_3$. This induces a mass for the tau
neutrino since $\nu_{\tau}$ mixes with the neutralinos (see the
Appendix). This mass turns out to be proportional to an {\sl
effective} neutralino mixing parameter $\xi \equiv (\mu v_3+\epsilon_3
v_1)^2$ characterizing the violation of R--parity, either through
$v_3$ or $\epsilon_3$. One can show \cite{DJV} that this parameter
corresponds to the R--parity violating VEV in the rotated basis. If we
stick to the simplest unified supergravity version of the model with
universal boundary conditions for the soft breaking parameters, then
$\xi$ will be small since contributions arising from {\sl gaugino}
mixing will cancel, to a large extent, those from {\sl Higgsino}
mixing.  This cancellation will happen automatically so that in this
case $m_{\nu_{\tau}}$ will be naturally small and radiatively 
calculable in terms of the bottom Yukawa coupling $h_b$, thus accounting 
naturally for the smallness of the $\nu_{\tau}$ mass in this model.  In the 
language of \cite{align} one can say that universality of the soft breaking 
terms implies an approximate and radiatively calculable {\sl
alignment} and as a result a suppression in $\xi$ and on $m_{\nu_{\tau}}$. 
This offers a {\sl hybrid} scenario combining the see-saw and radiative
schemes of $\nu$ mass generation.  The r\^ole of the right-handed mass
which appears in the see-saw model is played by the neutralino mass
(which lies at the SUSY scale) while the r\^ole of the seesaw-scheme
Dirac mass is played by the {\sl effective} neutralino mixing $\xi$
which is induced radiatively. The $\nu_{\tau}$ mass induced this way is
directly correlated with the magnitude of the effective parameter
$\xi$ so that R--parity violation acts as the origin for neutrino
mass.  In Fig.~\ref{mnt_xi_new} we display the allowed values of
$m_{\nu_{\tau}}$.
One can see that $m_{\nu_{\tau}}$ values can cover a very wide range,
from MeV values comparable to the present LEP limit \cite{ntbound}
down to values in the eV range, even though the individual values of
$v_3$ and $\epsilon_3$ can be rather large, e. g. $v_3, \epsilon_3
\sim$ 100 GeV, because they are naturally ``aligned'', without
any need for fine-tuning to get small $m_{\nu_{\tau}}$.  The alignment 
follows from the universality of the soft breaking parameters at the
unification scale \cite{epsrad}.  However, for generality, in this
paper we take the parameters at the weak scale independent (\ie\, no
universality assumption), and always impose $m_{\nu_{\tau}} \lsim 20$
MeV which corresponds to the laboratory limit.

This brings us to a discussion about cosmology. Clearly our model
leads to a tau--neutrino which can be much heavier than the limits
that follow from the cosmological critical density as well as
primordial nucleosynthesis would allow \cite{KT}.  However, in this
model the $\nu_{\tau}$ is unstable and decays via neutral current into three
lighter neutrinos \cite{2227}. In order for this mode to be efficient
we estimate that $m_{\nu_{\tau}}$ must exceed 100 keV or so. On the other 
hand, to avoid problems with primordial nucleosynthesis it is safer to 
consider masses below 1 MeV or so.  In order to sharpen these estimates
(which are not strict bounds) a detailed investigation is required
\cite{epscosmo}.

One should bear in mind, however, that $m_{\nu_{\tau}}$ can be as large as 
the present laboratory bound \cite{ntbound} in the more complete versions
of the model in which R--parity is broken spontaneously due to
sneutrino expectation values \cite{mv90,beyond}. This is so because
such models contain a majoron, denoted as J, which opens new decay
channels $\nu_{\tau}$ into $\nu + J$ where $\nu$ is a lighter neutrino
\cite{V} as well as new annihilation channels $\nu_{\tau}+\nu_{\tau}\to J+J$.
It has been shown explicitly that the lifetimes that can be achieved
in the spontaneous broken R--parity versions of the model can be
sufficiently short to obey the critical density limit
\cite{Romao92}. Moreover, it has been shown that the annihilation
channel is efficient enough in order to comply the primordial
nucleosynthesis bound \cite{DPRV}, while decays may also play an
important r\^ole \cite{unstable}.

Finally, a word about the magnitude of R--parity violation. It will
depend to some extent on the process considered. Some R--parity
violating observables turn out to be proportional to an {\sl
effective} neutralino mixing parameter $\xi \equiv (\mu v_3+\epsilon_3
v_1)^2$ characterizing the violation of R--parity, either through
$v_3$ or $\epsilon_3$. An example is the mass of the tau--neutrino
(see below).  However, not all R--parity-violating processes are
determined by $\xi$: some single production processes or R--parity
violating decay branching ratios, such as the ones discussed in the
present paper, depend separately on $v_3$ or $\epsilon_3$ and can be
rather large even for small $\xi$ and $m_{\nu_{\tau}}$. An obvious and 
important example is the decay of the lightest neutralino, which is 
determined by $\epsilon_3$ only since, in the rotated basis it is 
determined by the D Q L superpotential term only. A detailed discussion 
lies outside the scope of this paper \cite{DJV}.

\section{Scalar  Mass Matrices} 
 
The mass matrix of the charged scalar sector follows from the 
quadratic terms in the scalar potential 
\begin{equation} 
V_{quadratic}=[H_1^-,H_2^-,\tilde\tau_L^-,\tilde\tau_R^-] 
\bold{M_{S^{\pm}}^2}\left[\matrix{H_1^+ \cr H_2^+ \cr \tilde\tau_L^+ \cr 
\tilde\tau_R^+}\right]+... 
\label{eq:Vquadratic} 
\end{equation} 
For convenience reasons we will divide this $4\times4$ matrix into 
$2\times2$ blocks in the following way: 
\begin{equation} 
\bold{M_{S^{\pm}}^2}=\left[\matrix{ 
{\bold M_{HH}^2} & {\bold M_{H\tilde\tau}^{2T}} \cr 
{\bold M_{H\tilde\tau}^2} & {\bold M_{\tilde\tau\tilde\tau}^2} 
}\right] 
\label{eq:subdivM} 
\end{equation} 
where the charged Higgs block is 
\begin{eqnarray} 
&& {\bold M_{HH}^2}= 
\label{eq:subMHH} \\ \nonumber \\ 
&& \!\!\!\!\!\!\left[\matrix{ 
B\mu{{v_2}\over{v_1}}+\quarter g^2(v_2^2-v_3^2)+\mu\epsilon_3 
{{v_3}\over{v_1}}+\half h_{\tau}^2v_3^2+{{t_1}\over{v_1}} 
& B\mu+\quarter g^2v_1v_2 
\cr B\mu+\quarter g^2v_1v_2 
& B\mu{{v_1}\over{v_2}}+\quarter g^2(v_1^2+v_3^2)-B_2\epsilon_3 
{{v_3}\over{v_2}}+{{t_2}\over{v_2}} 
}\right] 
\nonumber 
\end{eqnarray} 
and $h_{\tau}$ is the tau Yukawa coupling. This matrix  
reduces to the usual charged Higgs mass matrix in the MSSM when we  
set $v_3=\epsilon_3=0$ and we call $m_{12}^2=B\mu$. The stau block is  
given by 
\begin{eqnarray} 
&& {\bold M_{\tilde\tau\tilde\tau}^2}= 
\label{eq:subtautau} \\ \nonumber \\ 
&& \!\!\!\!\!\!\!\!\!\!\!\!\!\!\left[\matrix{ 
\half h_{\tau}^2v_1^2-\quarter g^2(v_1^2-v_2^2)+\mu\epsilon_3 
{{v_1}\over{v_3}}-B_2\epsilon_3{{v_2}\over{v_3}}+{{t_3}\over{v_3}} 
& {1\over{\sqrt{2}}}h_{\tau}(A_{\tau}v_1-\mu v_2) 
\cr {1\over{\sqrt{2}}}h_{\tau}(A_{\tau}v_1-\mu v_2) 
& m_{R_3}^2+\half h_{\tau}^2(v_1^2+v_3^2) 
-\quarter g'^2(v_1^2-v_2^2+v_3^2) 
}\right] 
\nonumber 
\end{eqnarray} 
We recover the usual stau mass matrix again by replacing  
$v_3=\epsilon_3=0$, nevertheless, we need to replace the expression of the 
third tadpole in eq.~(\ref{eq:tadpoles}) before taking the limit. 
The mixing between the charged Higgs sector and the stau sector is 
given by the following $2\times2$ block: 
\begin{equation} 
{\bold M_{H\tilde\tau}^2}=\left[\matrix{ 
-\mu\epsilon_3-\half h_{\tau}^2v_1v_3+\quarter g^2v_1v_3 
& -B_2\epsilon_3+\quarter g^2v_2v_3 
\cr -{1\over{\sqrt{2}}}h_{\tau}(\epsilon_3v_2+A_{\tau}v_3) 
& -{1\over{\sqrt{2}}}h_{\tau}(\mu v_3+\epsilon_3v_1) 
}\right] 
\label{eq:subHtau} 
\end{equation} 
and as expected, this mixing vanishes in the limit $v_3=\epsilon_3=0$. 
The charged scalar mass matrix in eq.~(\ref{eq:subdivM}), 
after setting $t_1=t_2=t_3=0$, has determinant 
equal to zero since one of the eigenvectors corresponds to the charged 
Goldstone boson with zero eigenvalue. 
 
For completeness, we give the neutral Higgs sector mass matrices. The  
quadratic scalar potential includes 
\begin{equation} 
V_{quadratic}=\half[\varphi^0_1,\varphi^0_2,\tilde\nu_{\tau}^I] 
{\bold M^2_{P^0}}\left[\matrix{ 
\varphi^0_1 \cr \varphi^0_2 \cr \tilde\nu_{\tau}^I 
}\right] 
+\half[\chi^0_1,\chi^0_2,\tilde\nu_{\tau}^R] 
{\bold M^2_{S^0}}\left[\matrix{ 
\chi^0_1 \cr \chi^0_2 \cr \tilde\nu_{\tau}^R 
}\right]+... 
\label{eq:NeutScalLag} 
\end{equation} 
where the CP-odd neutral scalar mass matrix is 
\begin{equation} 
{\bold M^2_{P^0}}=\left[\matrix{ 
B\mu{{v_2}\over{v_1}}+\mu\epsilon_3{{v_3}\over{v_1}}+{{t_1}\over{v_1}} 
& B\mu & -\mu\epsilon_3 \cr B\mu & 
B\mu{{v_1}\over{v_2}}-B_2\epsilon_3{{v_3}\over{v_2}}+{{t_2}\over{v_2}} 
& -B_2\epsilon_3 \cr -\mu\epsilon_3 & -B_2\epsilon_3 &  
\mu\epsilon_3{{v_1}\over{v_3}}-B_2\epsilon_3{{v_2}\over{v_3}} 
+{{t_3}\over{v_3}} 
}\right] 
\label{eq:neuscaM} 
\end{equation} 
This matrix also has a vanishing determinant after the tadpoles are 
set to zero, and the zero eigenvalue corresponds to the mass of the 
neutral Goldstone boson. The usual MSSM mass matrix of the 
pseudoscalar Higgs sector is recovered in the limit $v_3=\epsilon_3=0$ 
in the upper--left $2\times2$ block, and the third component 
corresponding to the imaginary part of the sneutrino decouples from 
it. Note that in this limit, the MSSM mass of the sneutrino is 
recovered provided we replace the expression for the tadpole in 
eq.~(\ref{eq:tadpoles}) before taking the limit. 
 
The neutral CP-even scalar sector mass matrix in eq.~(\ref{eq:NeutScalLag}) 
is given by 
\footnote{Note that in the non-diagonal entries of this matrix the 
terms involving gauge couplings correct from those given in 
ref. \cite{eps0} by a factor 2. }  
\begin{eqnarray}  
&&{\bold M_{S^0}^2}= 
\label{eq:PseScalM}\\ \nonumber\\ 
&& \!\!\!\!\!\!\!\!\!\!\!\!\!\!\! 
\left[\matrix{ 
B\mu{{v_2}\over{v_1}}+\quarter g_Z^2v_1^2+\mu\epsilon_3 
{{v_3}\over{v_1}}+{{t_1}\over{v_1}}  
& -B\mu-\quarter g^2_Zv_1v_2  
& -\mu\epsilon_3+\quarter g^2_Zv_1v_3  
\cr -B\mu-\quarter g^2_Zv_1v_2  
& B\mu{{v_1}\over{v_2}}+\quarter g^2_Zv_2^2-B_2\epsilon_3 
{{v_3}\over{v_2}}+{{t_2}\over{v_2}}  
& B_2\epsilon_3-\quarter g^2_Zv_2v_3  
\cr -\mu\epsilon_3+\quarter g^2_Zv_1v_3  
& B_2\epsilon_3-\quarter g^2_Zv_2v_3  
& \mu\epsilon_3{{v_1}\over{v_3}}-B_2\epsilon_3{{v_2}\over{v_3}} 
+\quarter g^2_Zv_3^2+{{t_3}\over{v_3}}  
}\right] \nonumber 
\end{eqnarray} 
where we have defined $g_Z^2\equiv g^2+g'^2$. In the upper--left 
$2\times2$ block, in the limit $v_3=\epsilon_3=0$, the reader can 
recognize the MSSM mass matrix corresponding to the CP--even neutral 
Higgs sector. Similar to the previous case, in this limit the third 
component decouples from the other two and corresponds to the real 
part of the sneutrino, which become degenerate with the imaginary part 
of the sneutrino. Another way of looking at the separation of the 
sneutrino field into real and imaginary parts is through a 
sneutrino--anti-sneutrino 45 degrees mixing \cite{HowieSaSmix}. 
 
In the general case, there will be a mixing between the Higgs sector and 
the stau sector. The three mass matrices in eqs.~(\ref{eq:subdivM}),  
(\ref{eq:neuscaM}), and (\ref{eq:PseScalM}) are diagonalized by  
rotation matrices which define the eigenvectors 
\begin{equation} 
\left[\matrix{G^+ \cr H^+ \cr \tilde\tau_1^+ \cr \tilde\tau_2^+}\right]= 
{\bold R_{S^{\pm}}}\left[\matrix{ 
H_1^+ \cr H_2^+ \cr \tilde\tau_L^+ \cr \tilde\tau_R^+}\right] 
\,,\qquad\quad 
\left[\matrix{G^0 \cr A \cr \tilde\nu_{\tau}^I}\right]= 
{\bold R_{P^0}}\left[\matrix{ 
\varphi^0_1 \cr \varphi^0_2 \cr \tilde\nu_{\tau}^I}\right] 
\,,\qquad\quad 
\left[\matrix{h \cr H \cr \tilde\nu_{\tau}^R}\right]= 
{\bold R_{S^0}}\left[\matrix{ 
\chi^0_1 \cr \chi^0_2 \cr \tilde\nu_{\tau}^R}\right]\,, 
\label{eq:eigenvectors} 
\end{equation} 
and the eigenvalues 
$\rm{diag}(0,m_{H^{\pm}}^2,m_{\tilde\tau_1^{\pm}}^2, 
m_{\tilde\tau_2^{\pm}}^2)={\bold R_{S^{\pm}}}{\bold M_{S^{\pm}}^2} 
{\bold R_{S^{\pm}}^T}$ for the charged scalar sector, 
$\rm{diag}(0,m_A^2,m_{\tilde\nu_{\tau}^I}^2)={\bold R_{P^0}} 
{\bold M^2_{P^0}}{\bold R_{P^0}^T}$ for the CP--odd neutral scalar 
sector, and 
$\rm{diag}(m_h^2,m_H^2,m_{\tilde\nu_{\tau}^R}^2)={\bold R_{S^0}} 
{\bold M^2_{S^0}}{\bold R_{S^0}^T}$ for the CP--even neutral scalar 
sector. 
 
The labelling for the different eigenstates is as follows. In the 
charged scalar sector the eigenstate with zero mass is denoted 
$G^{\pm}$. Among the other three, we call staus the two eigenstates 
$S_i^{\pm}$ with the biggest stau component calculated with 
$(R_{S^{\pm}}^{i3})^2+(R_{S^{\pm}}^{i4})^2$, and by convention 
$m_{\tilde\tau^{\pm}_1}<m_{\tilde\tau^{\pm}_2}$. The remaining 
eigenstate is called charged Higgs $H^{\pm}$. In the neutral CP--odd 
sector, $G^0$ is the eigenstate $P^0_i$ with zero mass. Among the 
other two eigenstates, the one with largest stau component 
$(R_{P^0}^{i3})^2$ is called $\tilde\nu_{\tau}^I$, and the remaining 
state is the CP--odd Higgs $A$. Similarly, in the neutral CP--even 
sector, the state $S^0_i$ with the largest stau component 
$(R_{S^0}^{i3})^2$ is called sneutrino $\tilde\nu_{\tau}^R$. The other 
two are the neutral Higgs bosons $h$ and $H$, with $m_h<m_H$.  With 
this notation $H^{\pm}$ is the field which is mostly the MSSM charged 
Higgs, but with a small component of stau, and similarly for the 
neutral Higgs bosons. 
 
If a $3\times 3$ matrix ${\bold M}$ has a zero eigenvalue, then the 
other two eigenvalues satisfy 
\begin{equation} 
m_{\pm}={1\over 2}{\rm Tr}{\bold M} 
\pm {1\over2}\sqrt{\left({\rm Tr}{\bold M}\right)^2 
-4(M_{11}M_{22}-M_{12}^2+M_{11}M_{33}-M_{13}^2+M_{22}M_{33}-M_{23}^2)} 
\label{eq:EigenExact} 
\end{equation} 
The CP-odd neutral scalar mass matrix eq.~(\ref{eq:neuscaM}) has a zero 
determinant, so that its eigenvalues $m_{\tilde\nu_{\tau}^R}^2$ and 
$m_A^2$ can be calculated exactly with the previous formula.  The same 
can be done with the charged scalar mass matrix in the limit 
$h_{\tau}=0$.  In this case, the right stau decouples, and the 
eigenvalues $m_{H^{\pm}}^2$ and $m_{\tilde\tau_L^{\pm}}^2$ can be also 
calculated with eq.~(\ref{eq:EigenExact}).  Note that this limit is taken 
here only for the sake of illustration. In all numerical calculations 
we have used the realistic value for $h_{\tau}$ which is fixed through 
an exact tree-level relation eq.~(\ref{eq:htauf}) given in the Appendix. 
 
 
One can determine in the tree-level approximation the minimum of the 
scalar potential by imposing the condition of vanishing tadpoles in 
eq.~(\ref{eq:tadpoles}). One--loop corrections change these equations to 
\begin{equation} 
t_i=t_i^0-\delta t_i+T_i(Q) 
\label{eq:onelooptad} 
\end{equation} 
where $t_i$, with $i=1,2,3$, are the renormalized tadpoles, $t_i^0$ 
are the tree level tadpoles given in eq.~(\ref{eq:tadpoles}), $\delta t_i$ 
are the tadpole counter-terms, and $T_i(Q)$ are the sum of all 
one--loop contributions to the corresponding one--point functions with 
zero external momentum. The contribution from quarks and squarks to 
these tadpoles in our model can be found in ref.~\cite{epsrad}.  In an 
on shell scheme we identify the tree level tadpoles with the 
renormalized ones. Therefore, to find the correct minima we use 
eq.~(\ref{eq:tadpoles}) unchanged, where now all the parameters are 
understood to be renormalized quantities. 
 
We have used the diagrammatic (tadpole) method. Although equivalent to 
the effective potential method for minimization purposes 
\cite{diazhaberii}, the diagrammatic method is better than the  
effective potential when it comes to calculating the one--loop 
corrected scalar masses \cite{Vanderbilt}.  Following 
ref.~\cite{DiazHaberi} (see also ref.~\cite{chhothers}), we work in an 
on--shell scheme where by definition the tree level CP-odd Higgs mass 
$m_A$ and the tree level $W$--boson mass $m_W$ correspond to the 
respective pole masses.  In this case, the renormalized charged Higgs 
mass $m_{H^{\pm}}$ is equal to the tree level charged Higgs mass 
calculated in the previous section, plus the following one--loop 
contributions: 
\begin{equation} 
\Delta m_{H^{\pm}}^2={\rm Re}\left[A_{H^+H^-}(m_A^2+m_W^2)-A_{AA}(m_A^2) 
-A_{WW}(m_W^2)\right] 
\label{eq:RadCorrCH} 
\end{equation} 
where $A_{SS}(p^2)$, with $S=H^{\pm},A,W$, are self energies. Each 
self energy is infinite by itself, but $\Delta m_{H^{\pm}}^2$ is 
finite. 
 
Before turning to the numerical study of the charged Higgs mass
spectrum, let us mention that, throughout this paper, except the case
where the parameters have been fixed (Fig. 5 and 6), we have taken the
MSSM parameters varying in the range :
\begin{eqnarray} 
0.5 < & \tan\beta & < 90 \nonumber\\
0 \,{\mathrm GeV} < & M,M' & < 1000 \,{\mathrm GeV} \nonumber\\ 
0 \,{\mathrm GeV} < & m_{R_3}, m_{L_3} & < 300 \,{\mathrm GeV} \label{param}\\
-500 \,{\mathrm GeV} < & A_{\tau} & < 500 \,{\mathrm GeV} \nonumber\\ 
-200 \,{\mathrm GeV} < & \mu, B & < 0 \,{\mathrm GeV} \nonumber 
\end{eqnarray} 
and the two R--Parity violating parameters varying as:
\begin{eqnarray} 
-200 \,{\mathrm GeV} < & \epsilon_3 & < 200   \,{\mathrm GeV} \nonumber\\ 
-90 \,{\mathrm GeV} < & v_3 & < 90    \,{\mathrm GeV} \label{paramRV} 
\end{eqnarray} 
Note that $m^2_{H_1}$, $m^2_{H_2}$, and $B_2$ are fixed through the
tadpole equations given in eq.~(\ref{eq:tadpoles}). No big differences are
observed if we take $ \mu \geq 0 $ , and the sign of $ B $ is equal to
the sign of $ \mu $ because at the weak scale we have $ m_{A}^2 \propto
\mu B $.  We are interested in relatively light charged Higgs , so we
take $ | \mu | $ , $ | B | \leq 200 $ \,{\rm GeV} . Similarly we
are interested in relatively light staus , and that is why we take $
m_{R_{3}} $ , $ m_{L_{3}} $ $ \leq 300\: $ GeV.
 
\begin{figure} 
\centerline{\protect\hbox{\psfig{file=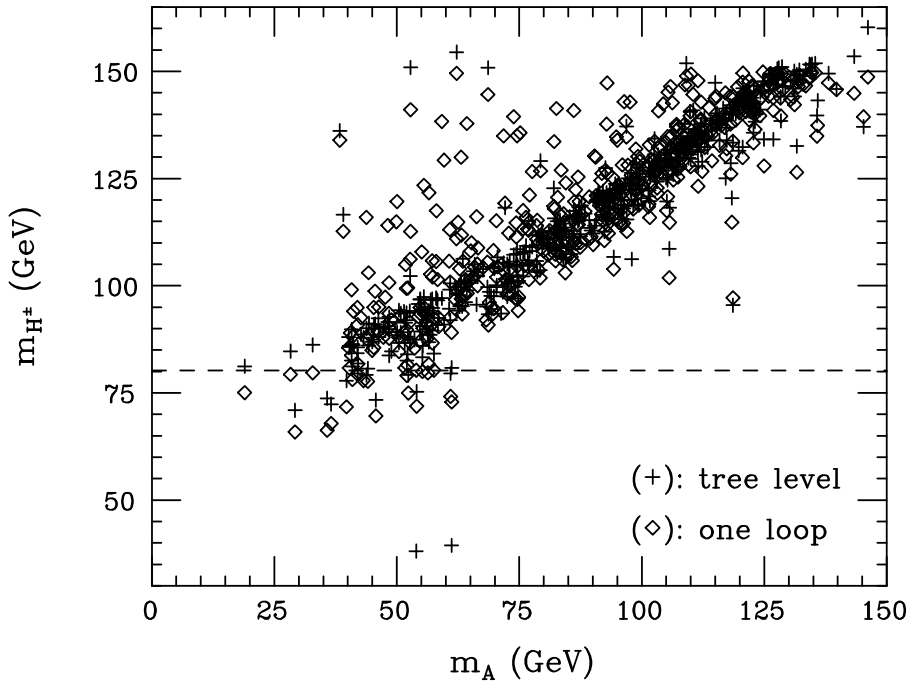,height=11cm,width=0.9\textwidth,angle=90}}} 
\caption{Tree level and one--loop charged Higgs boson mass  
as a function of the CP--odd Higgs mass $m_A$. The variation 
of parameters in the scan is indicated in the text.  The horizontal 
dashed line corresponds to the $W$-boson mass.} 
\label{fig:mchma} 
\end{figure}

We now turn to the numerical study of the lowest-lying charged scalar
boson mass. Our results are illustrated in Fig.~\ref{fig:mchma}.
In Fig.~\ref{fig:mchma} we display allowed values of charged Higgs 
boson mass as a function of the CP-odd neutral Higgs boson 
mass $m_A$. Here the main point to note is that $m_{H^{\pm}}$ can be 
lower than expected in the MSSM, even before including radiative 
corrections. This is due to negative contributions arising from the 
R-parity violating stau-Higgs mixing, controlled by the parameter 
$\epsilon_3$. We have varied the relevant model parameters in the range 
given by eq.~(\ref{param}) and eq.~(\ref{paramRV}).
 
The one loop correction $ \Delta m_{H^{\pm}}^2 $ in eq.~(\ref{eq:RadCorrCH}) 
depends on the soft squark masses $ M_{Q} $, $ M_{U} $ and $ M_{D} $,
and the tri-linear soft masses $ A_{t} $ and $ A_{b} $ . Since they appear
only through radiative corrections , for simplicity we have taken them
degenerate at the weak scale : $ M_{Q} = M_{U} = M_{D} = 1 \ {\mathrm TeV} $
and $ A_{t} = A_{b} = A_{\tau} $.
 
\begin{figure} 
\centerline{\protect\hbox{\psfig{file=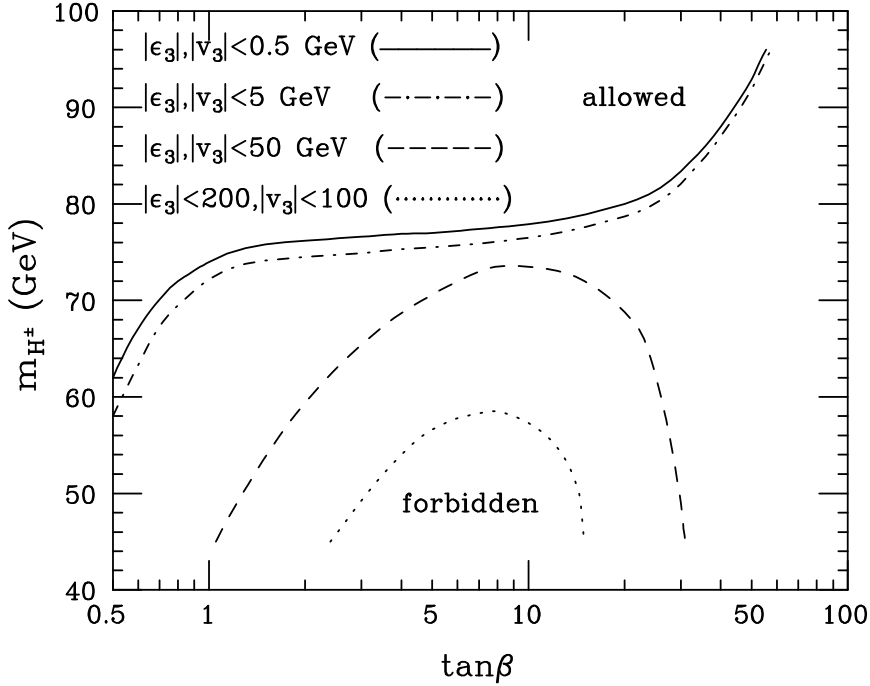,height=11cm,width=0.9\textwidth,angle=90}}} 
\caption{Minimum charged Higgs boson mass versus $\tan\beta$.
Each curve corresponds to a different range of variation of the
R--parity violating parameters $ \epsilon_{3} $ and $ v_{3} $}  
\label{fig:mh+tb} 
\end{figure}

An alternative way to display the influence of $\epsilon_3$ parameter 
on the charged Higgs boson mass can be seen in Fig.~\ref{fig:mh+tb}.
In this figure the curves corresponding to different $\epsilon_3$ and
$v_3$ values delimit the minimum theoretically allowed charged Higgs
boson mass corresponding to those specific values.  These curves are
found in a scan where the MSSM parameters are varied according to
eq.~(\ref{param}) and the R--parity violating parameters $\epsilon_{3}$ and 
$v_{3}$ are varied according to the label in Fig.~\ref{fig:mh+tb}.
Below each curve no points are found.  The radiatively corrected MSSM
prediction is recovered, as expected, when $\epsilon_3 = 0 =
v_3$. This is indicated by the solid line in the figure. For larger
values of the R-parity violating parameters one sees that the mass can
be substantially lower than the MSSM expectation. This is due, again,
to the negative contributions arising from the R-parity violating
stau-Higgs mixing, as discussed above.
 
\section{Production and Decays of Charged Scalars} 
 
Charged scalar pair production cross section can be calculated with
the aid of the $ZS_i^+S_j^-$ Feynman rule, which is equal to
$i\lambda_{ZS^+S^-}^{ij}(p+p')^{\mu}$ where $p$ and $p'$ are the
momenta in the direction of the positive electric charge flow. The
$\lambda$ couplings are equal to ${\bold\lambda}_{ZS^+S^-}={\bold
R_{S^{\pm}}} {\bold\lambda'_{ZS^+S^-}}{\bold R_{S^{\pm}}^T}$ where the
couplings in the unrotated basis are
\begin{equation} 
{\bold\lambda'_{ZS^+S^-}}={g\over{2c_W}}\left[\matrix{ -c_{2W} & 0 & 0 
& 0 \cr 0 & -c_{2W} & 0 & 0 \cr 0 & 0 & -c_{2W} & 0 \cr 0 & 0 & 0 & 
2s_W^2 }\right] \label{eq:ZSScouplings}  
\end{equation}  
The differential cross section for the production of two charged
scalars is
\begin{eqnarray} &&\!\!\!\!\!\!\!\!\!\!\!\! 
{{d\sigma}\over{d\cos\theta_{cm}}}(e^+e^-\rightarrow S_i^+S_j^-)= 
{1\over{32\pi s}}\lambda^{3/2}(1,m_{S^{\pm}_i}^2/s,m_{S^{\pm}_j}^2/s)
\sin^2\theta_{cm}
\label{eq:sigmaij}\\ &&\qquad \times\left[{{e^4}\over 
2}\delta_{ij}-{{ge^2}\over{2c_W}} 
g_V^e\lambda_{ZS^+S^-}^{ij}\delta_{ij} 
{s\over{s-m_Z^2}}+{{g^2}\over{8c_W^2}}(g_V^{e2}+g_A^{e2}) 
\lambda_{ZS^+S^-}^{ij2}{{s^2}\over{(s-m_Z^2)^2}}\right] \nonumber
\end{eqnarray}
where $g_V^e=\half-2s_W^2$ and $g_A^e=\half$ and 
$\lambda(a,b,c)=a^2+b^2+c^2-2ab-2ac-2bc$. 
\begin{figure} 
\centerline{\protect\hbox{\psfig{file=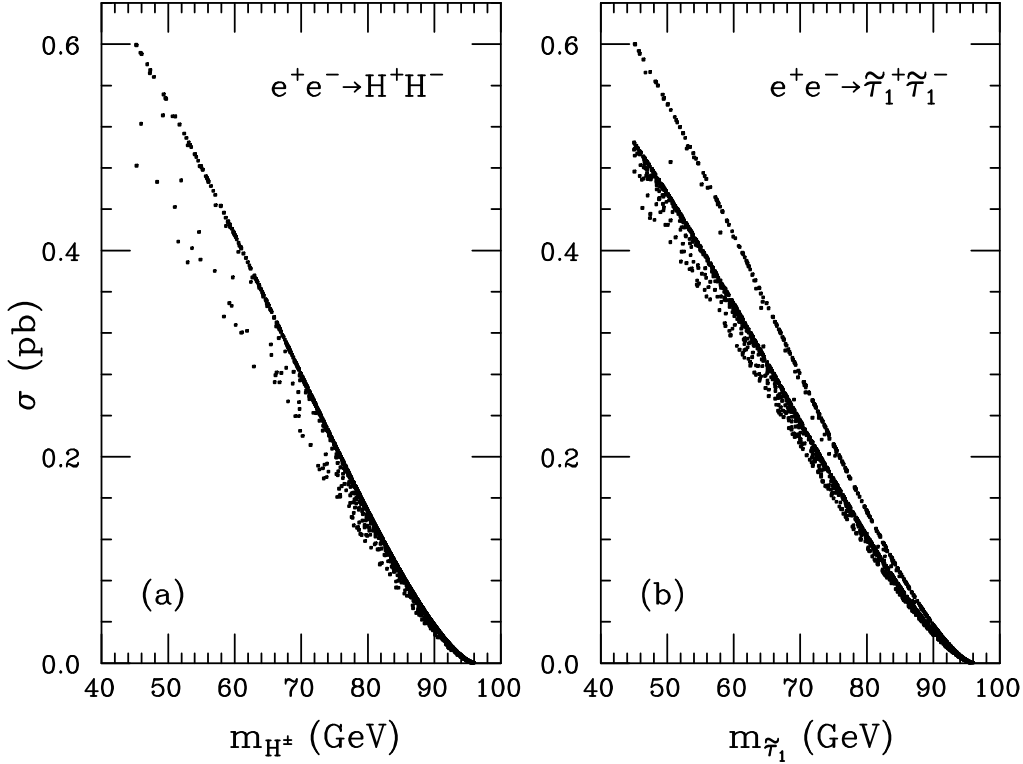,height=11cm,width=0.9\textwidth,angle=90}}} 
\caption{Total production cross section of a pair of (a) charged Higgs
bosons and (b) light staus, as a function of their mass. The centre of
mass energy is 192 GeV. }
\label{fig:xspair} 
\end{figure}  

In Fig.~\ref{fig:xspair} we plot the production cross section of a pair 
of charged scalars as a function of its mass, considering $\sqrt{s}=192$ 
GeV. In Fig.~\ref{fig:xspair}a we have $\sigma(e^+e^-\longrightarrow H^+H^-)$ 
and most of the points fall on the MSSM curve. The points that deviate 
from the main curve are due to mixing between charged Higgs and right stau. 
In fact, if the right stau were decoupled from the rest of the charged 
scalars, the charged Higgs pair production cross section would be  
identical to the MSSM for any value of $\epsilon_3$ or $v_3$, and the  
reason is that the upper--left $3\times 3$ relevant sub-matrix of  
$\bold\lambda'_{ZS^+S^-}$ in eq.~(\ref{eq:ZSScouplings}) is proportional  
to the identity. 
 
In Fig.~\ref{fig:xspair}b we have
$\sigma(e^+e^-\longrightarrow\tilde\tau_1^+\tilde\tau_1^-)$ as a
function of $m_{\tilde\tau_1^{\pm}}$. The points concentrate around
the two MSSM curves corresponding to the cases where
$\tilde\tau_1^{\pm}$ is mainly left stau (upper curve) and where
$\tilde\tau_1^{\pm}$ is mainly right stau (lower curve). The smallness
of the right-handed stau cross section relative to the left-handed one
is understandable because the right-handed stau is an $SU(2)_L$
singlet. Again, the points which deviate from these curves are due to
the mixing between the left and right staus.
 
\begin{figure} 
\centerline{\protect\hbox{\psfig{file=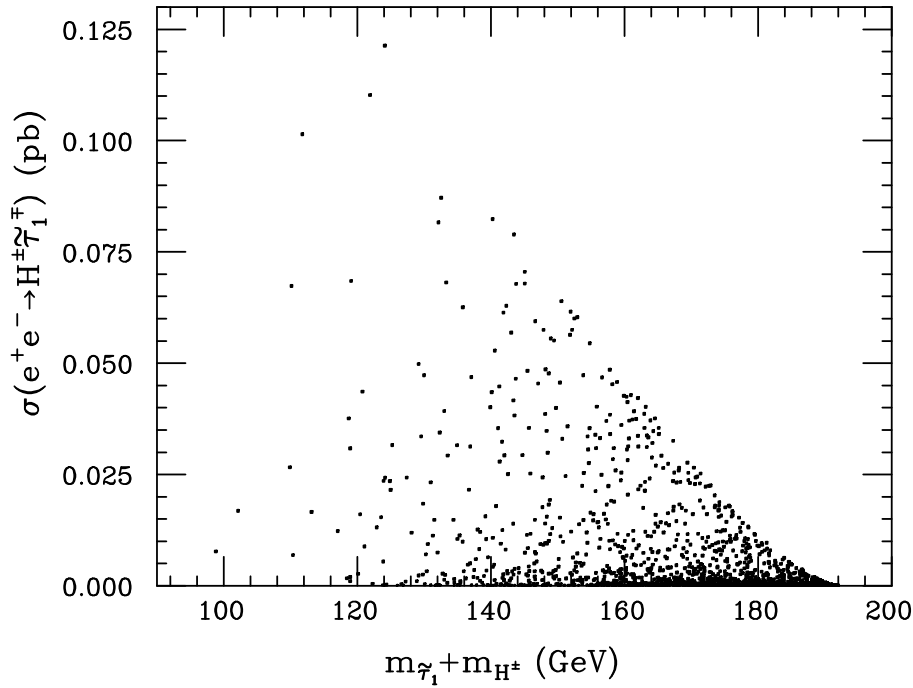,height=11cm,width=0.9\textwidth,angle=90}}} 
\caption{Mixed $H^{\pm}\tilde\tau_1^{\mp}$ production cross section 
as a function of $m_{H^{\pm}}+m_{\tilde\tau_1^{\mp}}$ at 192 GeV
centre-of-mass energy.} 
\label{fig:Smixed} 
\end{figure}  
An interesting characteristic of our model is the mixed production 
$e^+e^-\longrightarrow H^{\pm}\tilde\tau_1^{\mp}$ which is absent in  
the MSSM. In Fig.~\ref{fig:Smixed} we plot the total mixed production 
cross section, defined by  
$\sigma(e^+e^-\longrightarrow H^{\pm}\tilde\tau_1^{\mp})\equiv 
\sigma(e^+e^-\longrightarrow H^+\tilde\tau_1^-)+ 
\sigma(e^+e^-\longrightarrow H^-\tilde\tau_1^+)$, as a function of 
the sum of the final product masses $m_{H^{\pm}}+m_{\tilde\tau_1^{\pm}}$ 
for $\sqrt{s}=192$ GeV. This mixed production cross section can be sizable,  
with a maximum value of the order of 0.12 pb. 

As we have already seen, our model allows strong charged Higgs stau 
mixing, and this can substantially affect both the masses and the 
couplings. As a result the decay branching patterns of the charged 
scalar bosons can be significantly affected.  
 
\begin{figure} 
\centerline{\protect\hbox{\psfig{file=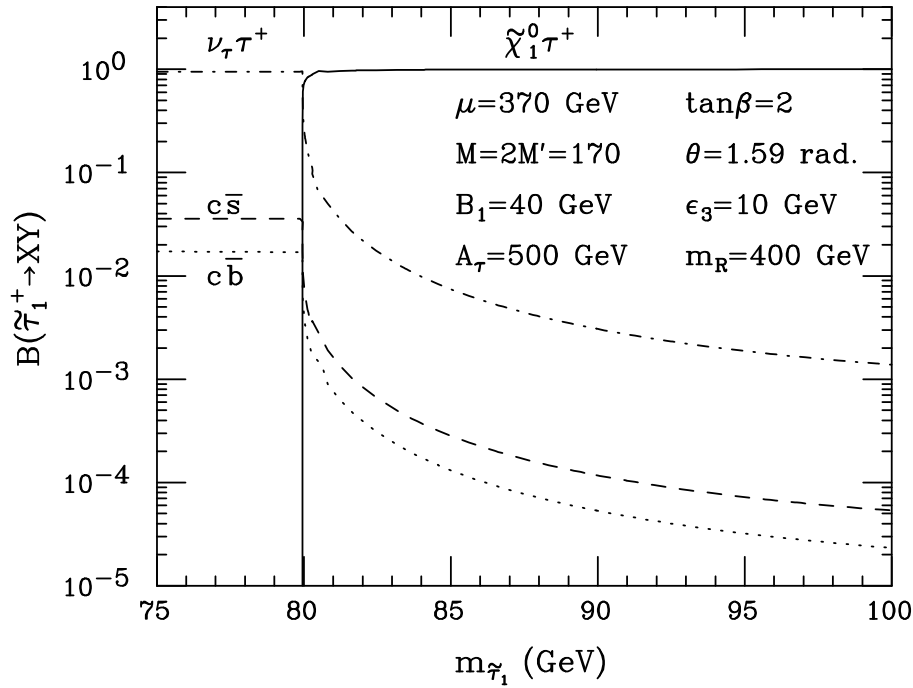,height=11cm,width=0.9\textwidth,angle=90}}} 
\caption{Stau branching ratios possible in our model for a particular choice
of parameters. Note the neutralino threshold below which only
R--parity violating decays are present.}  
\label{fig:v3stB} 
\end{figure}

We now turn to a discussion of the charged scalar boson decays.  Our 
first result here relates to the stau. In Fig.~\ref{fig:v3stB} we display 
the stau decay branching ratios below and past the neutralino 
threshold. In this figure we have fixed the parameters as  
\begin{eqnarray} 
\mu = 370 \ \,{\mathrm GeV} & \tan\beta = 2   \nonumber\\  
v_3 = -4.7 \ \,{\mathrm GeV} & M = 2M' = 170 \ {\mathrm GeV}  \nonumber\\ 
B = 40 \ \,{\mathrm GeV} & \epsilon_{3} = 10 \ {\mathrm GeV} \nonumber\\ 
A_{\tau} = 500 \ \,{\mathrm GeV}  & m_{R_3} = 400 \ \,{\mathrm GeV}  
\label{v3stB.} 
\end{eqnarray}
\noindent
in such a way as to ensure that the lightest neutralino is about 
80 GeV in mass and thus may be produced as a decay product of a stau 
produced at LEP II energies. 

In addition, we have chosen $ \epsilon_{3} = 10 \ \,{\mathrm GeV} $
and $ v_{3} = -4.7$ GeV ($\theta = 1.59$ rad) to demonstrate that we
don't need large R--parity violating parameters to obtain sizeable
effects. Otherwise, the choice of parameters is arbitrary .

Below the neutralino threshold, the stau is the LSP and will have only
R-parity violating decays, therefore totally un-suppressed, even for
the case of small R--parity breaking mixing.  The main modes of stau
decay in this case are into $\nu_{\tau}\tau$, $c\bar{s}$ and $c\bar{b}$, as
clearly seen from Fig.~\ref{fig:v3stB}. Moreover, one sees that for
typical values of the R--parity breaking parameters, the stau will
decay inside the detector.

\begin{figure} 
\centerline{\protect\hbox{\psfig{file=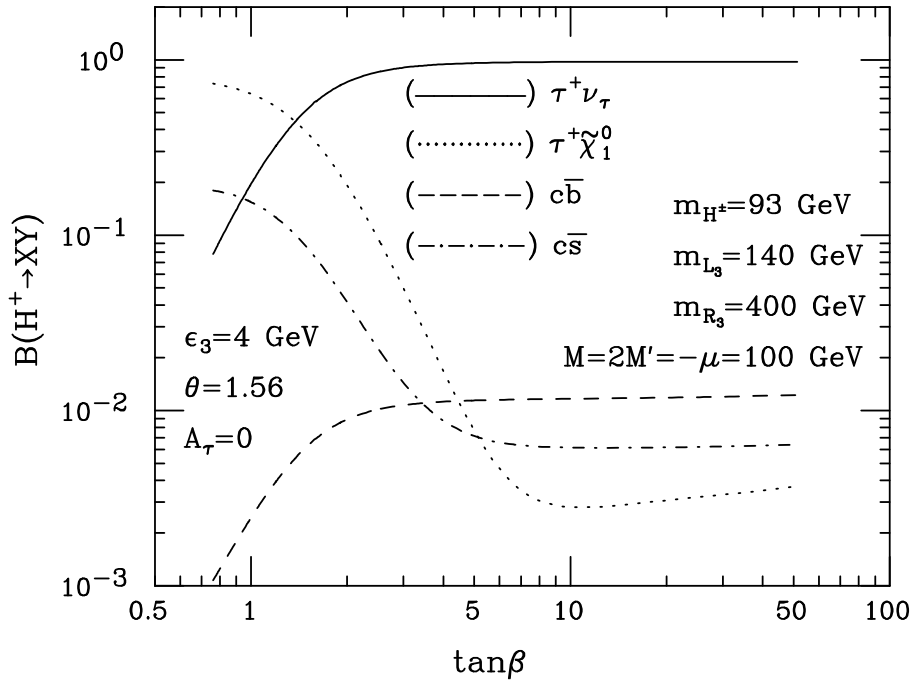,height=11cm,width=0.9\textwidth,angle=90}}} 
\caption{Charged Higgs branching ratios possible in our model for a
particular choice of parameters. The R--parity violating decay
dominates at low $\tan\beta$.}
\label{fig:h+br} 
\end{figure}

Finally, for the case of the R-parity violating charged Higgs boson 
decays one can see from Fig.~\ref{fig:h+br} that the branching ratios into 
supersymmetric channels can be comparable or even bigger than the 
R-parity conserving ones, even for relatively small values of 
$\epsilon$ and $v_3$.  

Indeed it is explicitly seen from Fig.~\ref{fig:h+br} that, in the
region of small $\tan\beta$, the R-parity violating Higgs boson decay
branching ratios can exceed the conventional ones and may reach values
close to 100 \%, since the R-parity--conserving decay $H^{\pm}\to
\tau\nu_{\tau}$ is proportional to $\tan^2\beta$ and so is usually 
dominant for larger $\tan\beta$.  This figure was obtained for a fixed
choice of parameters, given as
\begin{eqnarray} 
\mu = -100 \ \,{\mathrm GeV} &  m_{L_3} = 140 \ \mbox{GeV} \nonumber\\  
v_3 = 2.66 \mbox{GeV} & M = 2M' = 100 \ \mbox{GeV}  \nonumber\\ 
 m_{H^\pm} = 93 \ \mbox{GeV} & \epsilon_{3} = 4  \ \mbox{GeV} \nonumber\\ 
A_{\tau} = 0 \ \,\mbox{GeV}  & m_{R_3} = 400 \ \,\mbox{GeV}  
\label{rpv+br.} 
\end{eqnarray} 
Note that in eq.~(\ref{rpv+br.}) we have chosen $\mu$ with the opposite 
sign of eq.~(\ref{v3stB.}). 
This is done to see that the trends in Fig.~\ref{fig:v3stB} are  
not a peculiarity of that particular choice of parameters. 
Again we have taken small R--parity violating parameters:
$\epsilon_{3} = 4 \,{\mathrm GeV} $ and $v_3 = 2.66$ GeV. In 
order to get a ligth charged Higgs with $ m_{H^{\pm}} = 93 \,{\mathrm
GeV} $ we take $|\mu |$ and $|B|$ small.
 
We have also checked that a cosmologically safe 1 MeV $\nu_{\tau}$ would 
not modify appreciably our conclusions. For example, we have verified that
Fig. 7 remains unchanged if we vary $m_{\nu_{\tau}}$ up to 1 MeV.  The 
reason is that, although typically correlated with $m_{\nu_{\tau}}$ the 
R--parity violating branching ratio may be large if the Higgs and stau 
masses are relatively close to each other, even for much smaller 
$m_{\nu_{\tau}}$ values.

\begin{figure} 
\centerline{\protect\hbox{\psfig{file=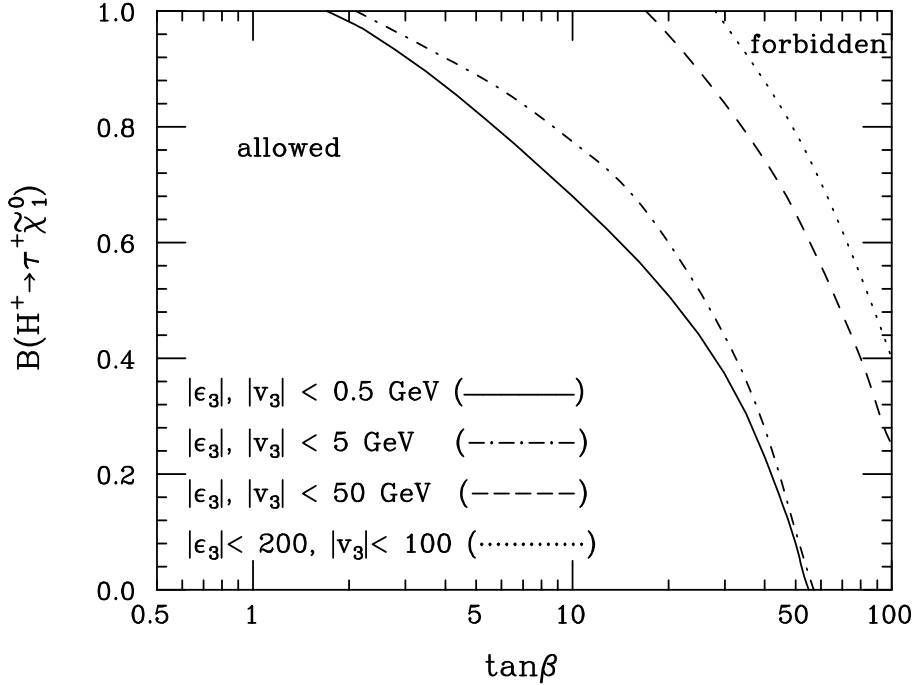,height=11cm,width=0.9\textwidth,angle=90}}} 
\caption{The curves denote the maximum attainable R-parity-violating
charged Higgs branching ratio versus $\tan\beta$. }
\label{rpv+br} 
\end{figure}

Another way to see that the dominance of R--parity-violating Higgs
boson decays is not an accident of the above parameter choice is
illustrated in Fig.~\ref{rpv+br}. The various curves denote the
maximum attainable values for the R--parity-violating Higgs boson
branching ratio $ B(H^{+} \longrightarrow \tau^{+}
\tilde\chi^{0}_{1})$.  The parameter space is scanned and the curves
represent the boundaries above which no points are found. The MSSM
parameters are varied according to eq.~(\ref{param}) and the range of
variation of the R--parity violating parameters $ \epsilon_{3} $ and
$v_{3}$ is indicated in the figure. In this way and in absolute
generality, we demonstrate that even for very small R--parity
violating parameters the branching ratio $B(H^{+} \longrightarrow
\tau^{+} \tilde\chi^{0}_{1}) $ can be close to unity, and that in the
region of $\tan\beta \gg 1 $ the decay $ H^{+} \longrightarrow
\tau^{+} \nu_{\tau} $ dominates.

\section{Discussion} 
 
We have seen in the last section [see Fig.~\ref{fig:v3stB}] that if the 
lightest stau $\tilde\tau_1^{\pm}$ is the LSP it will decay only through 
R-parity-violating interactions, to cs or $\tau\nu_{\tau}$. As a 
result it leads to decay signatures which are identical to those of 
the charged Higgs boson in the MSSM. However, if it is not the LSP the 
$\tilde\tau_1^{\pm}$ is more likely to have standard 
R-parity-conserving decays such as neutralino plus $\tau$, leading to 
signals that can be drastically different from those expected in the 
MSSM and which would arise from $\tilde\chi^0\longrightarrow 
\nu_{\tau}Z^*$ or $\tilde\chi^0\longrightarrow \tau W^*$. Unless 
$\epsilon_3$ and $v_3$ are extremely small, the neutralino will decay 
inside the detector. For the case of stau pair production in $e^+e^-$ 
colliders, such as LEP II, this would imply a plethora of new high 
fermion-multiplicity events (multi-jets and/or multi-leptons) . For 
example, di--tau + 4 jets + missing energy if both neutralinos decay 
into jets through neutral currents, or 4 taus + 4 jets if both 
neutralinos decay into jets through charged currents. Such processes 
are expected to have high rates and negligible background.  As for 
hadron colliders, we also can have very high leptonic multiplicity 
events such as six leptons of which at least two are taus, plus 
missing momentum. This should be easy to see at the LHC, due again to 
the negligible standard model background. For a recent discussion see, 
for example, ref.  \cite{gluino paper}. 
 
If the charged Higgs is the lightest charged scalar boson one can 
distinguish two scenarios. If the R-parity-violating parameter 
$\epsilon_3$ is small, the charged Higgs boson mainly decays to 
standard model fermions, thus conserving R-parity.  However ( as 
pointed out in ref. \cite{eps0} for the case of neutral Higgs) it is 
possible to obtain very large branching ratios for R-parity-violating 
charged Higgs boson decays, even for moderate $\epsilon_3$ values, as 
long as the mass difference between Higgs and staus is not too 
large. This happens because the R-parity-violating decay rates are 
governed by gauge strength interactions, whereas the 
R-parity-conserving ones are determined by Yukawa couplings. 

In the opposite case of large $\epsilon_3 \sim m_W$ we always 
expect large branching ratios for R-parity-violating charged Higgs 
boson decay modes. As a result, one expects again a large number of 
novel signatures arising from neutralino decays. These would be the 
same large multiplicity events that we mentioned above for stau pair 
production. 
 
Contrary to the MSSM, where stau and charged Higgs boson signal 
topologies are different, in our model they can be identical, for 
suitably chosen parameters. If one of these signals is observed, it 
will be difficult to know if it comes from a charged Higgs or a stau. 
To disentangle them, measurements of masses and decay rates of both 
particles would be required.  Alternatively, less experimental 
information would be required to separate the origin of the signals if 
the particle mass spectrum were predicted from theory, as happens in 
supergravity versions of this model \cite{epsrad}. 
 
Another interesting feature of our model is the mixed production 
$e^+e^-\longrightarrow H^{\pm}\tilde\tau^{\mp}$. If 
$m_{\chi^0_1}<m_{\tilde\tau_1^{\pm}}$ then one can produce interesting 
signatures like di-tau + di-jets + missing energy.  This is obtained 
when $H^{\pm}\longrightarrow \tau^{\pm}\nu_{\tau}$ and 
$\tilde\tau^{\mp}\longrightarrow\tau^{\mp}\tilde\chi^0_1 
\longrightarrow \tau^{\mp}q\overline{q}\nu_{\tau}$. Although the cross 
section for this case is typically smaller, one can see from 
Fig.~\ref{fig:Smixed} that for many choices of parameters it may be 
non-negligible. 
 
\section{Conclusion} 
 
In summary we have considered the most salient aspects of the 
phenomenology of the charged scalar boson sector in the simplest 
effective low-energy R-parity breaking model characterized by a 
bilinear violation of R-parity in the superpotential.  We have shown 
that the mass of the charged Higgs boson can be lower than expected in 
the MSSM, even before including radiative corrections. We have also 
studied the charged scalar boson decay branching ratios and show that 
the R-parity violating decay rates can be comparable or even bigger 
than the R-parity conserving ones. Moreover, if the stau is the LSP it 
will have only decays into standard model fermions, therefore totally 
un-suppressed. In the opposite case where it is heavier than the 
lightest neutralino one expects a plethora of exotic high 
fermion multiplicity events for which the standard model backgrounds 
should be rather small or absent. A detailed analysis of the 
detectability prospects of the related signatures at future 
accelerators lies outside the scope of the present paper and will be 
taken up elsewhere. 
 
\section*{Acknowledgements} 
 
This work was supported by DGICYT under grants PB95-1077 and by the 
TMR network grant ERBFMRXCT960090 of the European Union. M. A. D. was 
supported by a DGICYT postdoctoral grant, A. A. was supported by a 
CSIC-UK Royal Society postdoctoral grant, while J. F. and M. A. G-J 
were supported by Spanish MEC FPI fellowships. 
 
\section*{Appendix} 
 
For completeness, in this Appendix we collect the Feynman rules
relevant for the study of charged scalar decays into two fermions,
where these two fermions are a chargino-tau and a neutralino-neutrino.
We also give the exact formula, eq.~(\ref{eq:htauf}), for the tau-lepton
Yukawa coupling $h_{\tau}$ which differs in this model from that of
the MSSM.  First we set our conventions in the fermionic sector.
 
As we have already learned, the charged Higgs states $H_1^2$ and 
$H_2^1$ mix with the stau states $\tilde\tau_L^+$ and $\tilde\tau_R^+$ 
and form a set of four charged scalar eigenstates $S_i^{\pm}$ with 
$i=1,2,3,4$. The charged scalar mass matrix $\bold{M_{S^{\pm}}^2}$ in 
eq.~(\ref{eq:subdivM}) is diagonalized by a $4\times 4$ rotation 
matrix $\bold R_{S^{\pm}}$ defined in eq.~(\ref{eq:eigenvectors}). 
 
In a similar way, charginos mix with the tau lepton forming a set of  
three charge fermions $F_i^{\pm}$, $i=1,2,3$. In a basis where  
$\psi^{+T}=(-i\lambda^+,\widetilde H_2^1,\tau_R^+)$ 
and $\psi^{-T}=(-i\lambda^-,\widetilde H_1^2,\tau_L^-)$, the charged 
fermion mass terms in the lagrangian are 
\begin{equation} 
{\cal L}_m=-{1\over 2}(\psi^{+T},\psi^{-T}) 
\left(\matrix{{\bold 0} & \bold M_C^T \cr {\bold M_C} &  
{\bold 0} }\right) 
\left(\matrix{\psi^+ \cr \psi^-}\right)+h.c. 
\label{eq:chFmterm} 
\end{equation} 
where the chargino/tau mass matrix is given by 
\begin{equation} 
{\bold M_C}=\left[\matrix{ 
M & {\textstyle{1\over{\sqrt{2}}}}gv_2 & 0 \cr 
{\textstyle{1\over{\sqrt{2}}}}gv_1 & \mu &  
-{\textstyle{1\over{\sqrt{2}}}}h_{\tau}v_3 \cr 
{\textstyle{1\over{\sqrt{2}}}}gv_3 & -\epsilon_3 & 
{\textstyle{1\over{\sqrt{2}}}}h_{\tau}v_1}\right] 
\label{eq:ChaM6x6} 
\end{equation} 
and $M$ is the $SU(2)$ gaugino soft mass. 
We note that chargino sector decouples from the tau sector in the limit 
$\epsilon_3=v_3=0$. As in the MSSM, the chargino mass matrix is  
diagonalized by two rotation matrices $\bold U$ and $\bold V$ 
The tau Yukawa coupling $h_{\tau}$ is chosen such that one of the  
eigenvalues is equal to the tau mass. This is calculated from the  
vacuum expectation values of the model through an exact tree 
level relation given by 
\begin{equation} 
h_{\tau}^2=\frac{2M_\tau ^2}{v_1^2}\left( \frac{%
f+g(\varepsilon _3,v_3)}{f-\frac 2{v_1^2}h(\varepsilon _3,v_3)}\right)   
\label{eq:htauf} 
\end{equation} 
where :
$$ 
\quad f=g^2\left( \frac 12M_\tau ^2(v_1^2+v_2^2)+M\mu v_1v_2-\frac 
14g^2v_1^2v_2^2\right) +(\mu ^2-M_\tau ^2)(M_\tau ^2-M^2)  
$$ 
$$ 
\begin{array}{c} 
g(\varepsilon _3,v_3)=\frac 12v_3^2g^2\left( M_\tau ^2-\mu ^2-\frac 
12g^2v_2^2\right) -\varepsilon _3^2\left( M^2-M_\tau ^2+\frac 
12g^2v_1^2\right)  \\  
-\varepsilon _3v_3g^2(Mv_2+\mu v_1) 
\end{array} 
$$ 
$$ 
\begin{array}{c} 
h(\varepsilon _3,v_3)=\varepsilon _3v_3\left( \mu v_1(M_\tau ^2-M^2)+\frac 
12g^2Mv_1^2v_2\right) +\frac 12\varepsilon _3^2v_3^2\left( M^2-M_\tau 
^2\right)  \\  
+\frac 12v_3^2\left( M_\tau ^2(M_\tau ^2-M^2)-g^2M_\tau ^2(v_1^2+\frac 
12v_2^2)-g^2M\mu v_1v_2+\frac 12g^4v_1^2v_2^2\right)  \\  
+\frac 12\varepsilon _3v_3^3g^2Mv_2-\frac 14v_3^4g^2(M_\tau ^2-\frac 
12g^2v_2^2) 
\end{array} 
$$ 
In our model, the tau neutrino aquires mass, and this is due to a 
mixing between the neutralino sector and the neutrino--tau, forming a 
set of five neutral fermion $F^0_i$, $i=1,...5$. In the basis 
$\psi^{0T}= 
(-i\lambda',-i\lambda^3,\widetilde{H}_1^1,\widetilde{H}_2^2,\nu_{\tau})$ 
the neutral fermions mass terms in the lagrangian are given by 
\begin{equation} 
{\cal L}_m=-\frac 12(\psi^0)^T{\bold M}_N\psi^0+h.c.   
\label{eq:NeuMLag} 
\end{equation} 
where the neutralino/neutrino mass matrix is 
\begin{equation} 
{\bold M}_N=\left[  
\begin{array}{ccccc}  
M^{\prime } & 0 & -\frac 12g^{\prime }v_1 & \frac 12g^{\prime }v_2 & -\frac  
12g^{\prime }v_3 \\   
0 & M & \frac 12gv_1 & -\frac 12gv_2 & \frac 12gv_3 \\   
-\frac 12g^{\prime }v_1 & \frac 12gv_1 & 0 & -\mu  & 0 \\   
\frac 12g^{\prime }v_2 & -\frac 12gv_2 & -\mu  & 0 & \epsilon _3 \\   
-\frac 12g^{\prime }v_3 & \frac 12gv_3 & 0 & \epsilon _3 & 0  
\end{array}  
\right] 
\label{eq:NeuM5x5} 
\end{equation} 
and $M'$ is the $U(1)$ gaugino soft mass. This neutralino/neutrino mass  
matrix is diagonalized by a $5\times 5$ rotation matrix $\bold N$ such that 
\begin{equation} 
{\bold N}^*{\bold M}_N{\bold N}^{-1}={\rm diag}(m_{\chi^0_1},m_{\chi^0_2}, 
m_{\chi^0_3},m_{\chi^0_4},m_{\nu_{\tau}}) 
\label{eq:NeuMdiag} 
\end{equation} 
where by definition the eigenstate $F_5^0$ is the neutrino--tau, i.e., 
with the largest tau component $(N_{i5})^2$.  
 
Now we are ready to work out the $S^{\pm}_iF^{\mp}_jF^0_k$ Feynman rules.  
These vertices are denoted by 
\begin{equation} 
S^+_iF^-_jF^0_k\longrightarrow 
i\lambda _{S^{+}F^{-}F^0}^{L\,\,ijk}P_L+ 
i\lambda _{S^{+}F^{-}F^0}^{R\,\,ijk}P_R 
\label{eq:SFFvertex} 
\end{equation} 
where $P_L=\half(1-\gamma_5)$ and $P_R=\half(1+\gamma_5)$ are the usual 
left and right proyection operators. For simplicity, we work  
in a basis where the chaged scalars are unrotated. In  
this basis $S'^+=(H_1^{2*},H_2^1,\tilde\tau_L^+,\tilde\tau_R^+)$ the 
vertices are denoted by 
\begin{equation} 
S'^+_iF^-_jF^0_k\longrightarrow 
i\lambda'^{L\,\,ijk}_{S^{+}F^{-}F^0}P_L+ 
i\lambda'^{R\,\,ijk}_{S^{+}F^{-}F^0}P_R 
\label{eq:SpFFvertex} 
\end{equation} 
The relation between $\lambda$ and $\lambda'$ is given by 
\begin{equation} 
\lambda^{L\,\,ijk}_{S^{+}F^{-}F^0}={\bf R}_{S^{\pm }}^{il} 
\lambda'^{L\,\,ljk}_{S^{+}F^{-}F^0}\,,\qquad 
\lambda^{R\,\,ijk}_{S^{+}F^{-}F^0}={\bf R}_{S^{\pm }}^{il} 
\lambda'^{R\,\,ljk}_{S^{+}F^{-}F^0} 
\label{eq:LamLamp} 
\end{equation} 
and each of the $\lambda'$ can be read from the following Feynman rules 
\begin{eqnarray} 
H_1^{2*}F^-_iF^0_j&\longrightarrow&i\frac g2\left[-U_{i1}^*N_{j3}^*+ 
\frac 1{\sqrt{2}}U_{i2}^*\left(N_{j2}^*+\frac{g'}gN_{j1}^*\right)\right]  
(1-\gamma _5)+i\frac{h_{\tau}}2N_{j5}V_{i3}(1+\gamma _5)   
\nonumber\\ 
H_2^1F^-_iF^0_j&\longrightarrow&-i\frac g2\left[V_{i1}N_{j4}+ 
\frac 1{\sqrt{2}}V_{i2}\left(N_{j2}+\frac{g'}gN_{j1}\right)\right] 
(1+\gamma _5)   
\label{eq:SFFrules}\\ 
\tilde\tau_L^+F^-_iF^0_j&\longrightarrow&-i\frac g2\left[ 
U_{i1}^*N_{j5}^*-\frac 1{\sqrt{2}}U_{i3}^*\left(  
N_{j2}^*+\frac{g'}gN_{j1}^*\right)\right](1-\gamma _5)- 
i\frac{h_{\tau}}2N_{j3}V_{i3}(1+\gamma _5)   
\nonumber\\ 
\tilde\tau_R^+F^-_iF^0_j&\longrightarrow&-i\frac{g'}{\sqrt{2}}V_{i3} 
N_{j1}(1+\gamma _5)+i\frac{h_{\tau}}2\left(N_{j5}^*U_{i2}^*- 
N_{j3}^*U_{i3}^*\right)(1-\gamma _5) 
\nonumber 
\end{eqnarray} 
The reader can check that from eq.~(\ref{eq:SFFrules}) we can recover  
the MSSM Feynman rules taking the appropriate limits.

 \end{document}